\documentclass[
aps,%
11pt,%
final,
notitlepage,%
oneside,%
onecolumn,%
nobibnotes,%
nofootinbib,%
preprintnumbers,%
superscriptaddress,%
showpacs,%
centertags,%
showkeys,%
amsmath,%
amssymb]{revtex4}

\usepackage{graphicx} 
\usepackage{epsfig}
\usepackage{dcolumn} 

\def\phi{\varphi}

\begin{document}

\title{{Physics with $ep$ collisions at highest $Q^2$ and P$_{\rm T}$ at the
HERA collider.}
}

\author{Juraj Bracinik, for the H1 and ZEUS collaborations}
\email{bracinik@mppmu.mpg.de}
\affiliation{ MPI for Physics, Munich, Germany   }

\date{\today}

\begin{abstract}
The HERA collider with its center of mass energy of $318$ GeV makes it possible to study a wide range of electroweak physics as well as to search for physics beyond the Standard Model (SM).
In this article, recent results, obtained by the two collider experiments H1 and ZEUS, are reviewed. The cross sections for inclusive neutral current and charged current processes are shown, and results from a combined electroweak and QCD analysis of the data are discussed. Selected results from searches for physics beyond the SM are presented.

\end{abstract}


\maketitle

\section{Introduction}

HERA was the first and up to now the only $ep$ collider. It was built in Hamburg in Northern Germany and during most of its lifetime collided leptons with energy of $27.6$ GeV with protons with energy of $920$ GeV. Since the early $90$'s untill the end of June $2007$ it has delivered around $0.7$ fb$^{-1}$ of high quality data, almost equally shared between $e^+$ and $e^-$ beams and with longitudinal polarizations of different sign.
Inclusive Deep Inelastic Scattering (DIS) processes are for a given center-of-mass energy 
$\sqrt{s}$ characterized by two independent kinematic invariants. Commonly used variables are $Q^2$, defined as the virtuality squared of the exchanged boson, the Bjorken $x$ variable, in leading order equal to the fractional  proton momentum carried by the struck parton, and the inelasticity $y$, which in the proton rest frame is equal to the fractional energy loss of the beam lepton.  Only two of them are independent  as $Q^2$=$xys$.

\section{Basic features of Neutral Current (NC) and Charged Current (CC) processes at HERA}

There are two basic types of DIS processes at HERA. In the case of NC scattering, the scattered lepton is observed together with one or several 
hadron jets which balance the transverse momentum
(P$_{\rm T}$) of the lepton. In the case of CC scattering there is no scattered lepton observed in the final state. Therefore the CC events are characterized by significant missing transverse momentum since only one or several hadron jets are observed.

\begin{figure}[hhh]
\center
\epsfig{file=nc_cc_dsdq2_cteq6d.eps.gv,width=0.4\textwidth}
\epsfig{file=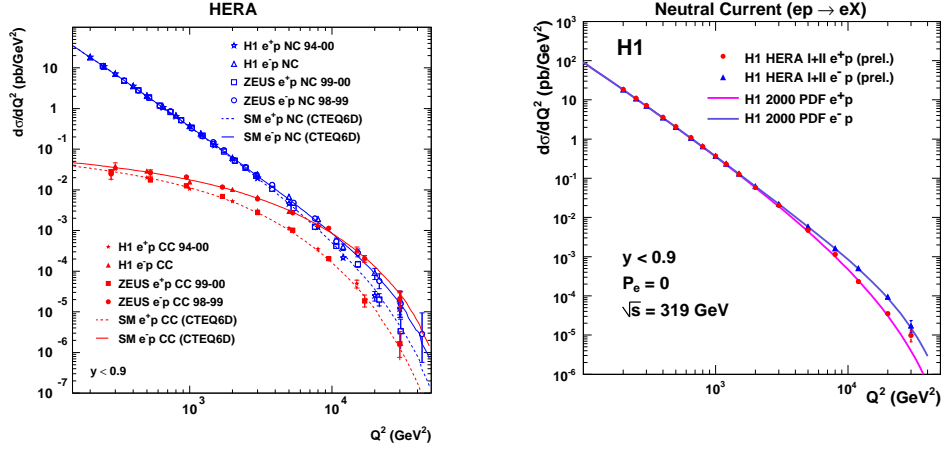,width=0.38\textwidth}
\caption{Differential cross section of NC and CC processes in $e^+p$ and 
$e^-p$ interactions as a function of $Q^2$ 
(HERA I data (left) and  full HERA I+II data set (right)). }
\label{fig:nc_cc_unpolarized} 
\end{figure}

The NC processes are  due to the exchange of the 
photon or 
the Z-boson between the lepton and the proton. Good knowledge of the electroweak vertex and the fact that neutral current processes in the phase space accessible by HERA are dominated by the photon exchange make it convenient to write the differential cross section in the following form~\cite{physrew}:
\begin{equation}
\frac{d^2 \sigma^{e^\pm p}_{\rm NC}}{dx dQ^2}=
\frac{2 \pi \alpha^2}{x Q^4} \left[  Y_+ \tilde{F_2} \mp Y_- 
x \tilde{F_3} - y^2  \tilde{F_L}  \right],
\label{eq:nc_section}
\end{equation}
where $Y_{\pm} = 1 \pm (1-y)^2$. The structure of the proton is parameterized by  three generalized structure functions $\tilde{F_2}$, $x \tilde{F_3}$ and $\tilde{F_L}$. The dominant contribution over the HERA phase space is given by  $F_2$, which in quark-parton model (QPM) is the charge square weighted sum of the quark and anti-quark parton distribution functions (pdfs). The generalized structure function  $x \tilde{F_3}$ starts to play a role only at high $Q^2$.
It is dominated by $\gamma-Z$ interference contribution and 
in the QPM it  is proportional to the pdf's of the valence quarks $2u_v+d_v$. 
The longitudinal structure function $ \tilde{F_L}$ contributes at high inelasticities and in the QPM it is equal to zero.

The CC processes are  due to  W exchange. Their cross section can be written in the form
\begin{equation}
\frac{d^2 \sigma^{e^\pm p}_{\rm CC}}{dx dQ^2}=
\frac{G_F^2}{4 \pi x} \left[  \frac{M_W^2}{M_W^2+Q^2}  \right]^2
\left[  Y_+W_2 \mp Y_- x W_3 - y^2 W_L
\right].
\label{eq:cc_section}
\end{equation}
In the QPM, the cross section for CC positron-proton scattering at large $x$
is sensitive mainly to the $d$-quark pdf, as $\sigma_{\rm CC}^{e^+p} \sim x[(\overline{u}+\overline{c}) +
(1-y)^2(d+s)]$, while for electron-proton scattering it depends primarily on
$u$-quark pdf, as
 $\sigma_{\rm CC}^{e^-p} \sim x[(u+c) + (1-y)^2 (\overline{d}+\overline{s})]$.

The phase space of HERA is determined by its center of mass energy
$\sqrt{s} =318$ GeV, sufficient to observe the electroweak effects both in NC and CC scattering.
As an example, the differential cross section of NC and CC scattering is shown in Fig.~\ref{fig:nc_cc_unpolarized} (left) as a function of $Q^2$. While at low $Q^2$ the cross sections of NC and CC scattering differ by orders of magnitude due to different propagators, at $Q^2$ comparable to the W mass squared, they become similar, providing a demonstration of electroweak unification. The CC cross section for $e^-p$ is larger than the one for $e^+p$, reflecting the fact that the proton consists mainly of $u$-type quarks.
The NC cross section as a function of $Q^2$ is also shown in Fig.~\ref{fig:nc_cc_unpolarized} (right) for two lepton beam charges, this time for the full HERA statistics. At high $Q^2$ the 
two cross sections start to differ. This charge asymmetry is described by the generalized structure function $x \tilde{F_3}$.

\section{Effect of longitudinal lepton beam polarization on NC and CC cross sections}
\begin{figure}[hhh]
\center
\epsfig{file=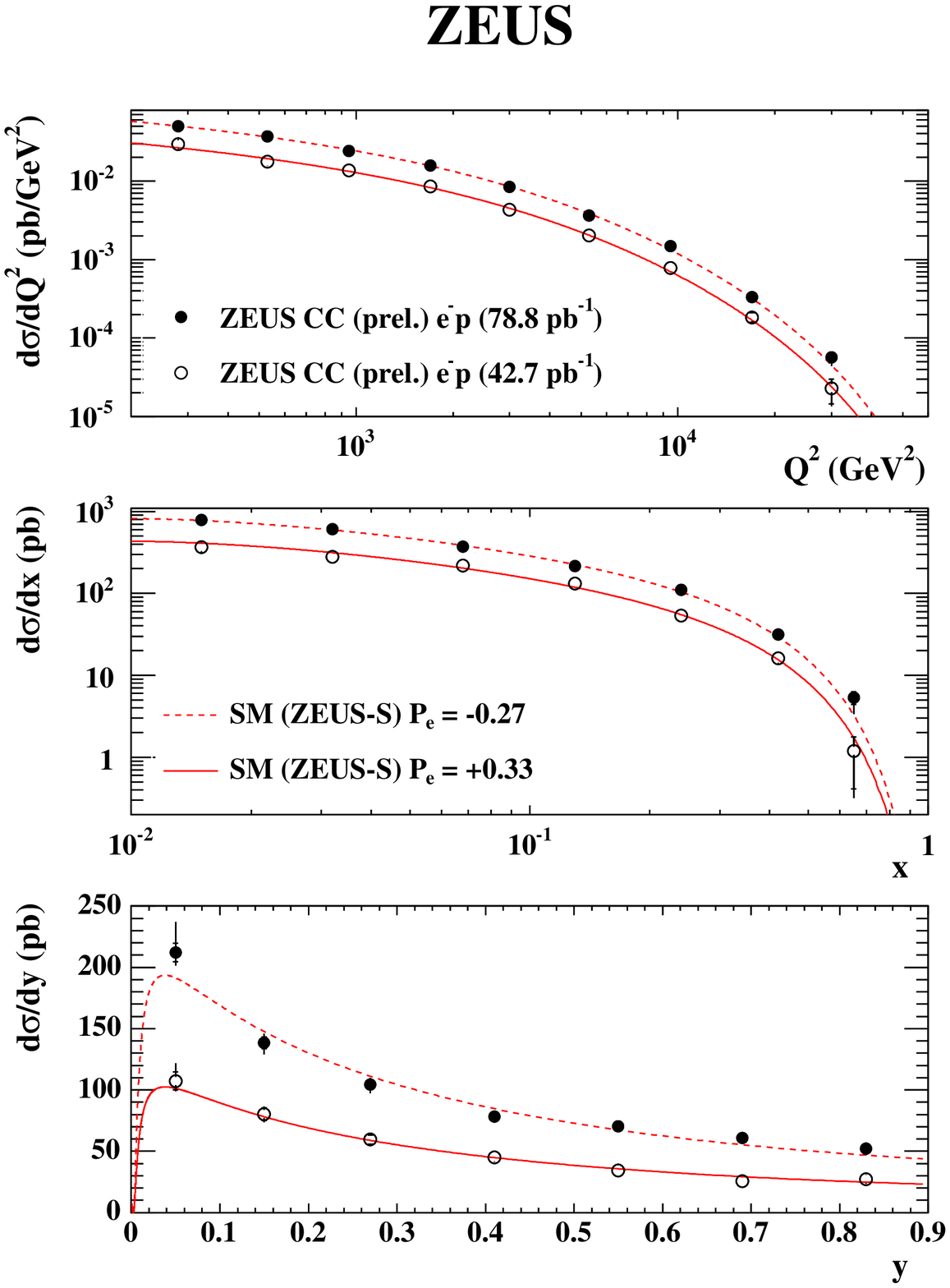,width=.4\textwidth}
\epsfig{file=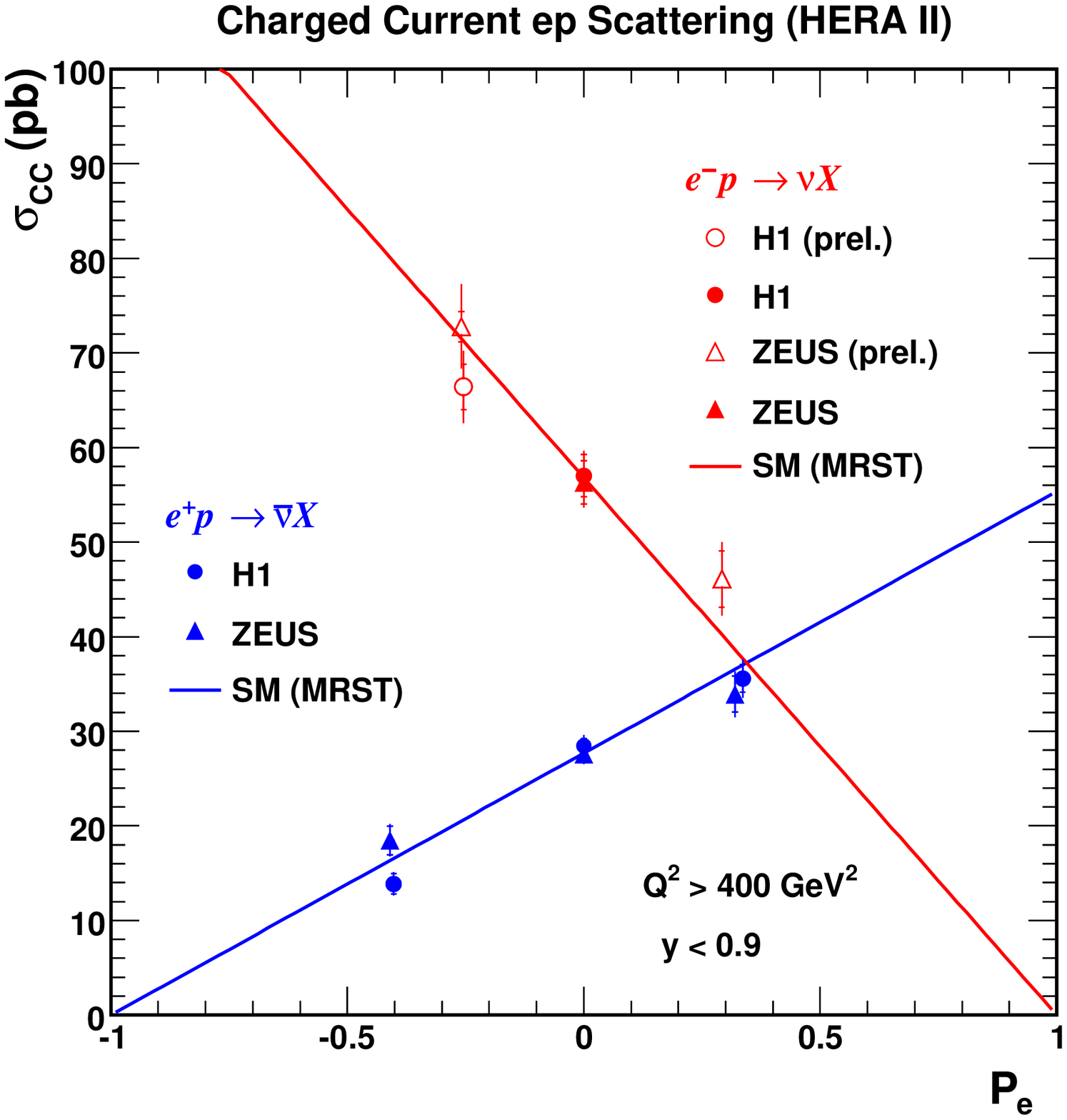,width=.5\textwidth}
\caption{Differential cross section of CC processes for $e^-p$
interactions as a function of
$Q^2$, $x$ and $y$ for right and left-handed longitudinal polarizations (left) and
the total cross section for $e^{\pm}p$ CC processes as a function of polarization (right).}
\label{fig:cc_diff_pol} 
\end{figure}
Since the HERA and detector upgrades in the years 2000-2002 (HERA II), it was possible to study  electroweak effects using longitudinally polarized lepton beams. The transverse polarization of the lepton beam builds up naturally due to the Sokolov-Ternov effect. During the upgrade spin rotators were installed that flip the polarization to longitudinal just before the interaction regions and flip it back after them.
The level of polarization is defined as $P=(N_R-N_L)/(N_R+N_L)$, where $N_R$ ($N_L$) is the number of right handed (left handed) leptons in a beam. The typical levels of polarizations delivered were 30-40 \%.          

The longitudinal polarization has a particularly striking effect on the CC cross section. In the Standard Model (SM) only left-handed particles (and right-handed anti-particles) interact via CC, and a linear dependence of the cross section on the polarization is expected, i.e. $\sigma_{\rm CC}^{e^{\pm}p}(P)= [1 \pm P] \sigma_{\rm CC}(P=0)$. Indeed, in 
Fig.~\ref{fig:cc_diff_pol} (left) one can see the measured differential CC cross section as a function of $Q^2$, $x$ and $y$ for two values of the right and left-handed longitudinal polarization. The differential cross sections for different polarizations have the same shape, but different normalization.
The total CC cross section for the two beam charges, integrated over the visible phase space defined by $Q^2>$ $400$ GeV$^2$ and $y<0.9$, is shown in 
Fig.~\ref{fig:cc_diff_pol} (right) as a function of $P$. The data are within current errors 
 in excellent agreement with the SM prediction, exhibiting a linear dependence on $P$.
Hypothetical right-handed CC, absent in the SM, would lead to a linear dependence of $\sigma_{\rm CC}^{e^\pm p}$ on $P$ with a nonzero intercept at the polarization 
$P( \mp 1)$. That is why the polarization dependence of $\sigma_{CC}$
is fit by a straight line and $\sigma_{CC}$ is  extrapolated to the helicity, where the cross section is expected to vanish according to the SM. For the case of the electron beam the extrapolation leads to 
$\sigma_{\rm CC}(e^-p, P_e=+1) = -0.9 \pm 2.9  ({\rm stat}) \pm 1.9  ({\rm sys}) \pm 2.9
  ({\rm pol})$ 
(H1) and $\sigma_{\rm CC}(e^-p, P_e=+1) = 0.8 \pm 3.1 ({\rm stat}) \pm 5.0  
({\rm sys+pol})$
(ZEUS). For the positron beam $\sigma_{\rm CC}(e^+p, P_e=-1)= 
-3.9 \pm 2.3  ({\rm stat}) \pm 0.7  ({\rm sys}) \pm 0.8  ({\rm pol}) $ (H1) and
 $\sigma_{\rm CC}(e^+p, P_e=-1)= 7.4 \pm 3.9  ({\rm stat}) \pm 1.2  
({\rm sys+pol})$.
Assuming the same couplings for the left-handed and hypothetical right-handed CC, these extrapolated cross sections are converted to 95 \% confidence limits on the mass of the heavy $W_{\rm R}$ boson, yielding $M(W_R)>208$ GeV (H1, $e^+p$),
$M(W_R)>186$ GeV (H1, $e^-p$) and $M(W_R)>180$ GeV (ZEUS, $e^-p$).
\begin{figure}[hhh]
\center
\epsfig{file=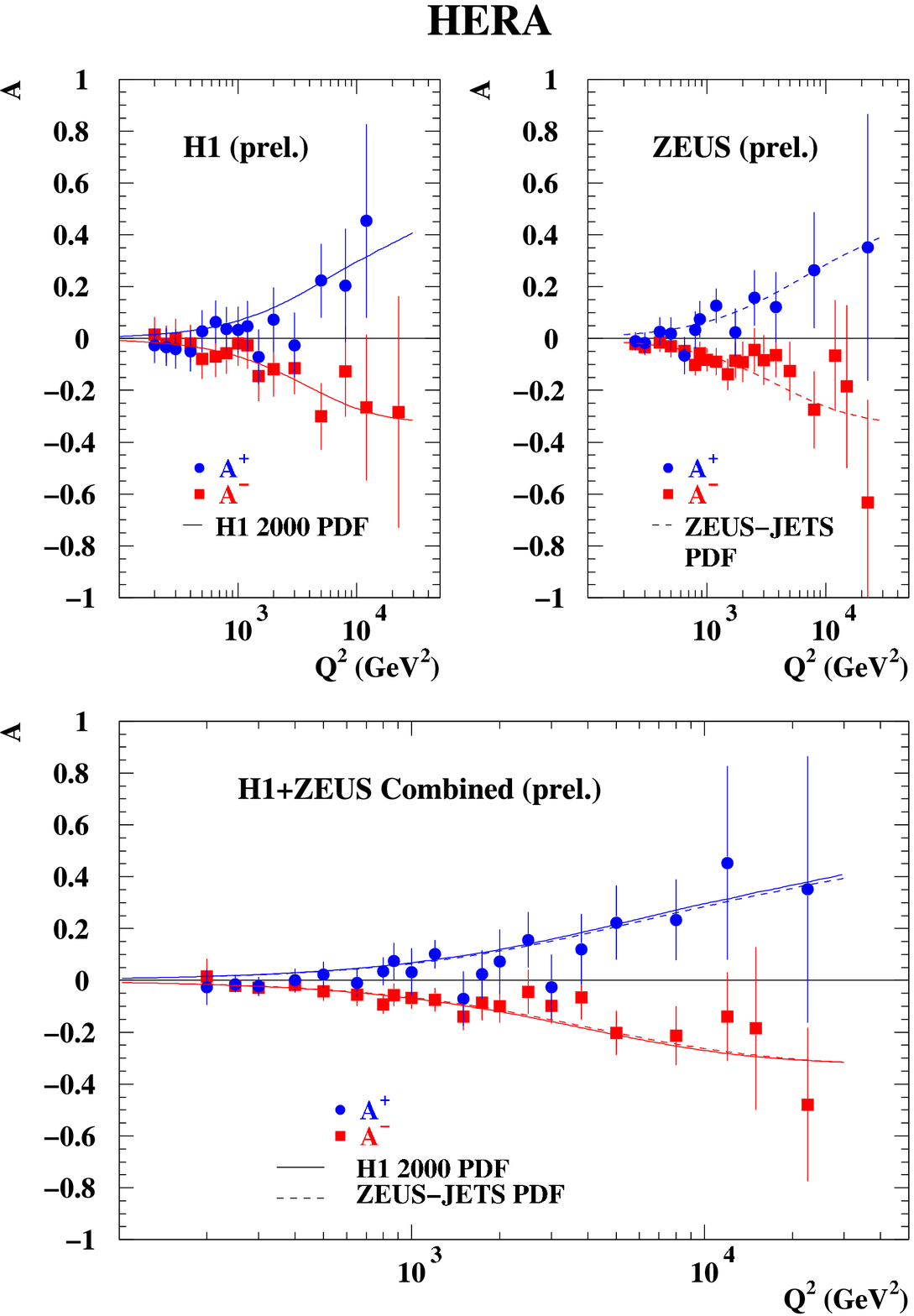,width=.435\textwidth}
\epsfig{file=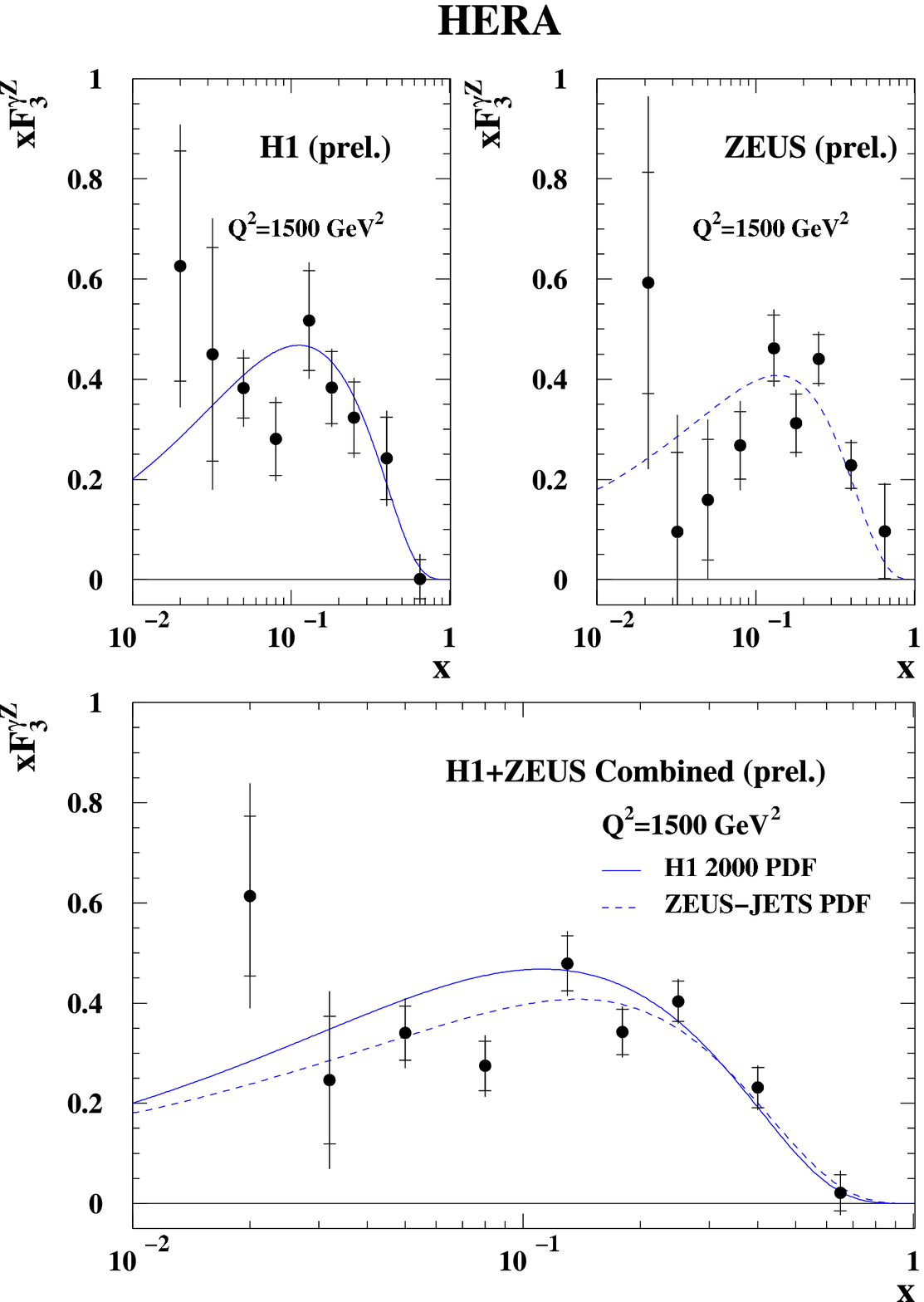,width=.435\textwidth}
\caption{The charge dependent polarization asymmetry $A^{\pm}$ as seen in NC scattering as a function of $Q^2$ (left) and the structure function $xF_3^{\gamma Z}$ as a function of $x$ (right). }
\label{fig:nc_pol} 
\end{figure}

The dependence of the NC cross section on the polarization is 
still linear but more evolved, as both generalized structure functions $\tilde{F_2}$ and $x\tilde{F_3}$ depend on the polarization.
The structure function $\tilde{F_2}$ consists of contributions from pure $\gamma$-exchange, pure $Z$-exchange and their interference. The contribution from pure $Z$-exchange is suppressed by a propagator term, and can be neglected, 
allowing to write
\(
\tilde{F_2} \approx F_2 - (v_e \pm P a_e) \chi_Z F_2^{\gamma Z},
\label{eq:f2_pol}
\)
where $\chi_Z = (1/sin^2 2\theta_W)(Q^2/(M_Z^2+Q^2))$, $v_e$ and $a_e$ are the vector and axial vector couplings of the electron to $Z$, and $\theta_W$ is the Weinberg angle. While the $\gamma$-exchange term is dominant, it does not depend on polarization. The vector coupling $v_e$ is an order of magnitude smaller then $a_e$ ($a_e=-0.5$, $v_e \approx -0.04$) and with good precision $\tilde{F_2} \approx F_2 \pm P a_e \chi_Z F_2^{\gamma Z}$.
In Fig.~\ref{fig:nc_pol} (left) the charge dependent polarization asymmetry $A^{\pm}$ is plotted as a function of $Q^2$. The charge dependent polarization asymmetry depends on the lepton beam charge and is defined as
$A^{\pm} = 2/(P_R-P_L) ((\sigma(e^{\pm},P_R) - \sigma(e^{\pm},P_L))/(\sigma(e^{\pm},P_R) + \sigma(e^{\pm},P_L))$. As seen in Fig.~\ref{fig:nc_pol} (left), the polarization asymmetry has the same size, but opposite sign for the two lepton beam charges and is well described by the SM prediction. The observation of the polarization asymmetry in NC scattering demonstrates parity violation at distances of $10^{-18}$ m.

The structure function $x \tilde{F_3}$ consists of contributions from pure $Z$-exchange and $\gamma$-$Z$ interference. As in the case of  $\tilde{F_2}$,  pure $Z$-exchange is supressed by the propagator and 
\(
x\tilde{F_3} \approx -(a_e \pm P v_e) \chi_Z x F_3^{\gamma Z}.
\label{eq:xf3_pol}
\)
Due to the  small value of $v_e$, $x\tilde{F_3}$ does not, to first order, depend on the polarization.
That is the reason why polarized NC cross sections can be used to measure the unpolarized structure function $x\tilde{F_3}$. The $e^+p$ and $e^-p$ data are corrected for polarization effects, then their difference is used to calculate $x\tilde{F_3}$. In Fig.~\ref{fig:nc_pol} (right), one can see the interference term $xF_3^{\gamma Z}$ as a function of $x$, extracted from the measured $x\tilde{F_3}$. This quantity provides a direct measurement of the valence quark distributions at $x$ values down to $10^{-2}$.

\section{The combined Electroweak and QCD analysis of NC and CC data}

\begin{figure}[hhh]
\center
\epsfig{file=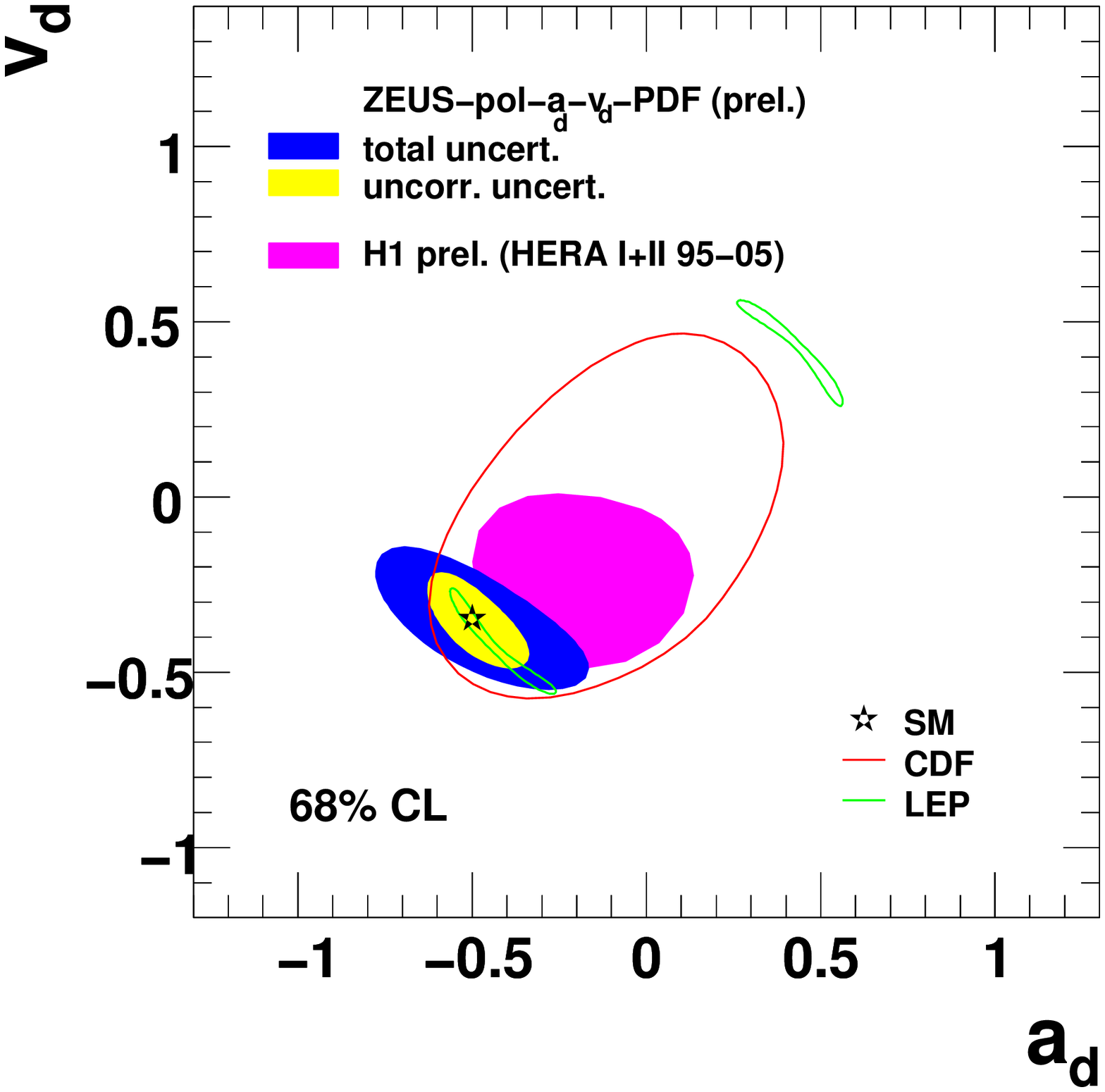,width=.45\textwidth}
\epsfig{file=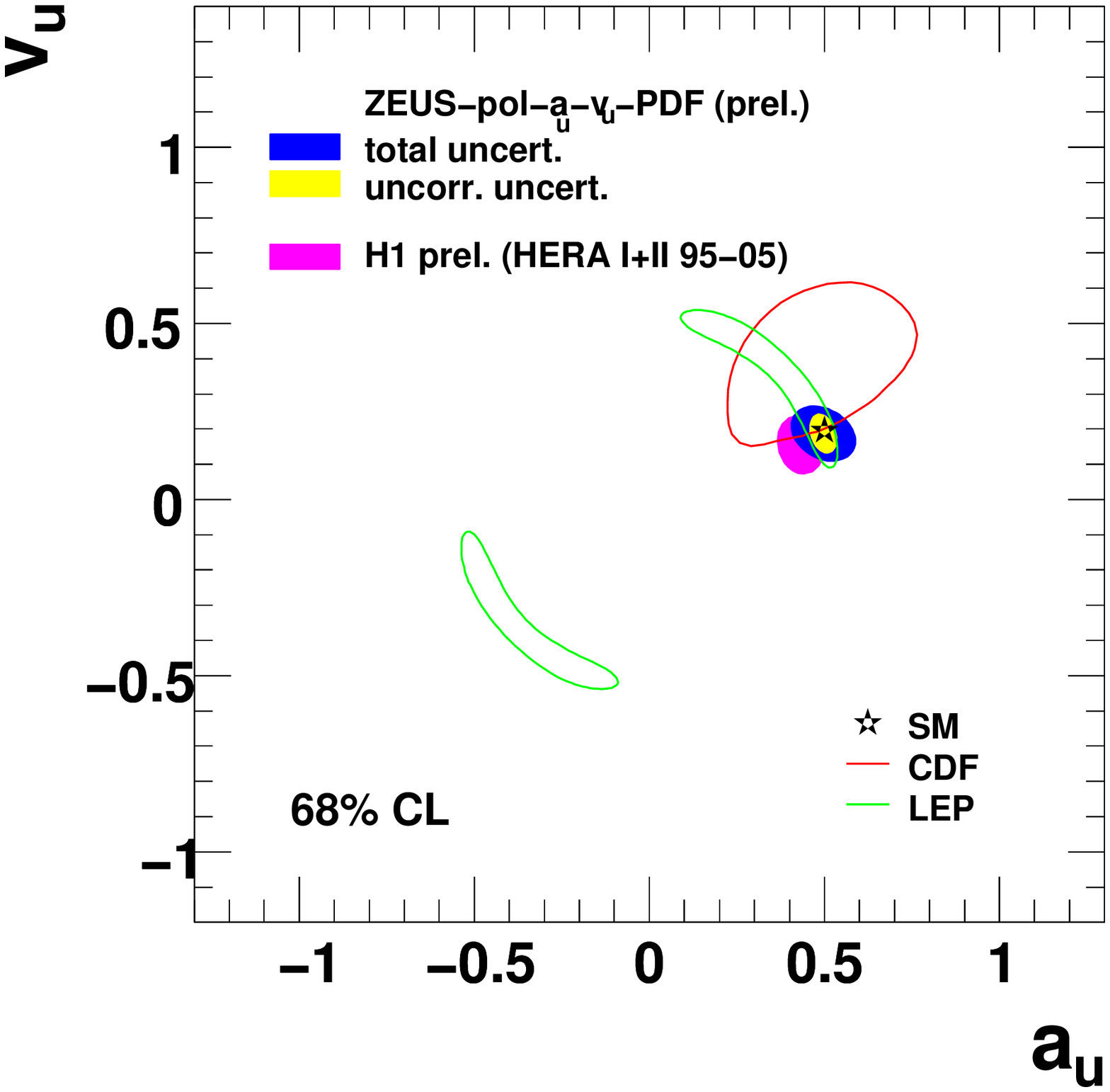,width=.45\textwidth}
\caption{The values of axial ($a_i$) and vector ($v_i$) couplings of light quarks to the $Z$-boson. The results were obtained by a combined Electroweak and QCD fit, determining the couplings of one type of quark while fixing the couplings for the other type of light quarks to the SM value. }
\label{fig:ew_couplings} 
\end{figure}
\begin{figure}[hhh]
\center
\epsfig{file=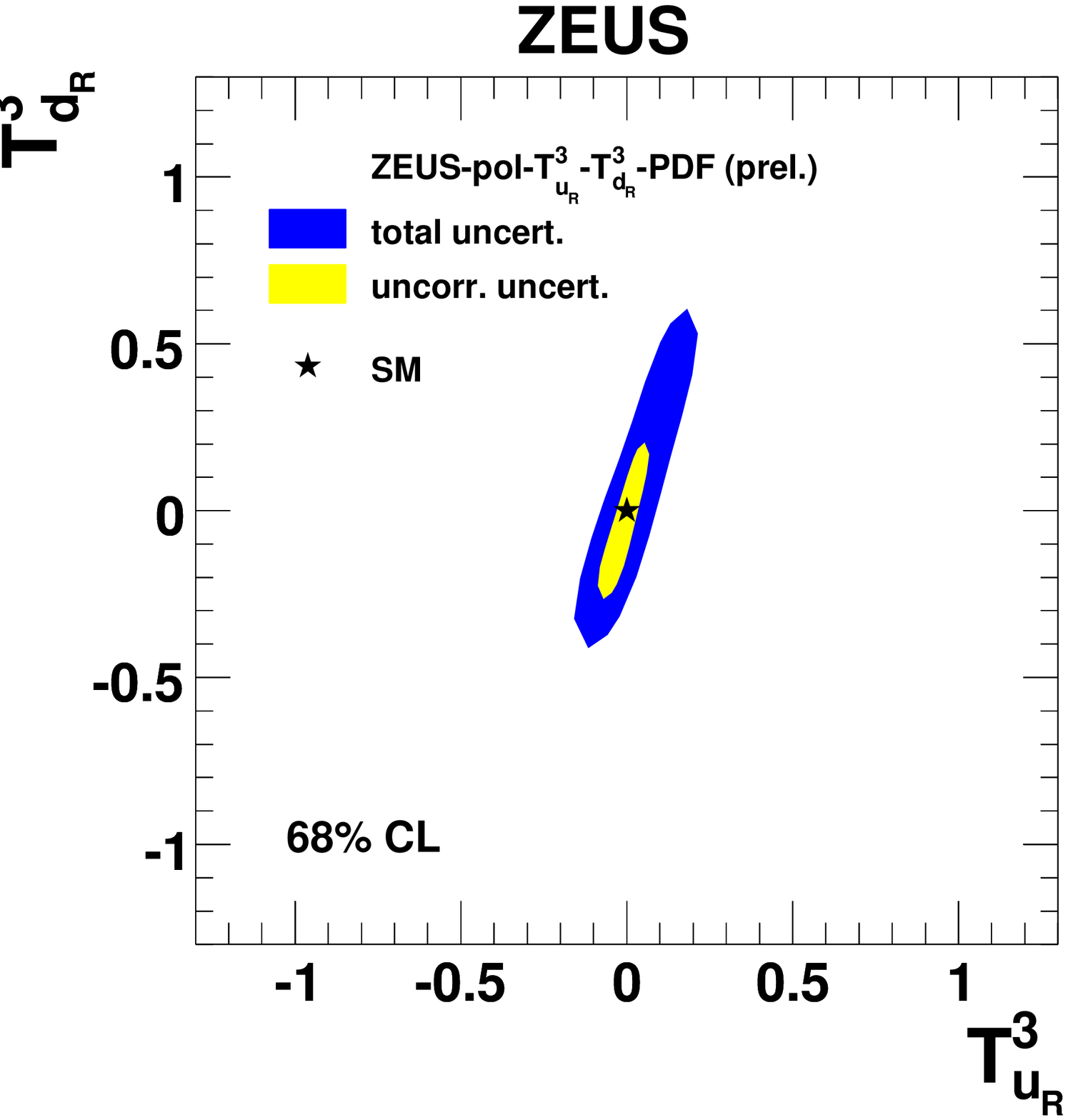,width=.45\textwidth}
\epsfig{file=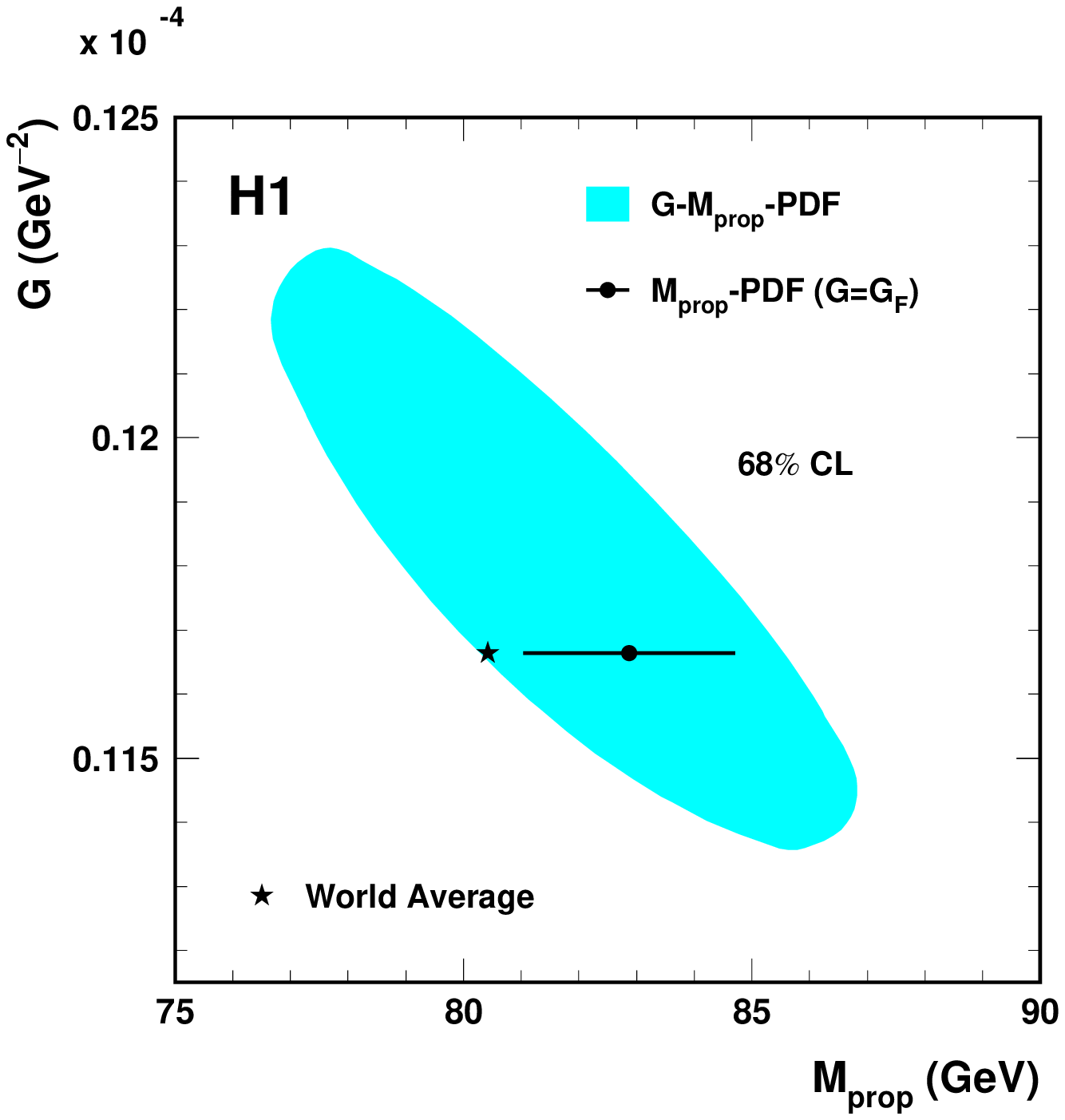,width=.5\textwidth}
\caption{The confidence limits on the right handed weak isospin (left) and the propagator mass of the object exchanged in CC events (right). }
\label{fig:ew_isospin_mass} 
\end{figure}

The precise and well understood data sets, collected by both HERA collider experiments, allow to perform an analysis in the framework of the SM. In such an analysis several data sets are used (NC and CC cross section, in the case of the ZEUS analysis also jets in photoproduction) to extract simultaneously the QCD (pdfs) and electroweak parameters.
While more details about the extracted pdfs can be found in~\cite{amanda}, here the extraction of electroweak parameters will be discussed.

The data from $ep$ collisions provide an interesting possibility to measure the vector and axial vector  couplings of light quarks to the $Z$-boson as both generalized structure functions $\tilde{F_2}$ and $x\tilde{F_3}$ depend on them. Indeed, 
in QPM $F_2^{\gamma Z} = 2 x \sum e_i v_i [q_i+\overline{q_i}]$, 
$F_2^{Z} =  x \sum (v_i^2 + a_i^2) [q_i+\overline{q_i}]$,
$xF_3^{\gamma Z} = 2 x \sum e_i a_i [q_i-\overline{q_i}]$ and
$xF_3^{Z} = 2 x \sum v_i a_i [q_i-\overline{q_i}]$, where $v_i$ and $a_i$ are the vector and axial vector couplings of quarks to the $Z$-boson, $e_i$ the quark charges and $q_i$ ($\overline{q_i}$) the quark (antiquark) pdfs. 
Due to the interference terms $F_2^{\gamma Z}$ and $xF_3^{\gamma Z}$ there is no sign ambiguity for $v_i$ and $a_i$ as there is the case in analysis of $e^+e^-$ annihilation data. The longitudinal polarization improves mainly the measurement of $v_i$.
The values of $v_i$ and $a_i$ resulting from a joint electroweak and QCD parameter fit
 are shown in Fig.~\ref{fig:ew_couplings}. 
The results, shown in the Fig.~\ref{fig:ew_couplings} are obtained by fitting the couplings
of $u$-type quarks ($d$-type quarks)  while fixing the couplings of $d$-type quarks ($u$-type quarks) to their SM value. As the proton consists mainly of $u$-type quarks, the fit is sensitive mainly to $u$ couplings, providing their best measurement available.

The measurement of vector and axial vector couplings can be used to set limits on the right handed weak isospin. The couplings are expressed in the form $a_q = T^3_{q,L}-T^3_{q,R}$ and
$v_q=T^3_{q,L}+T^3_{q,R} - 2e_q \sin^2 \theta_{W}$. The confidence limits on the  right handed isospin for $u$ and $d$ type quarks, obtained by fixing $T^3_{q,L}$ and $\sin^2 \theta_W$ to their SM values can be seen in 
Fig.~\ref{fig:ew_isospin_mass} (left). The results are in good agreement with the SM prediction of right handed isospin being equal to zero.

The $Q^2$ dependence of the CC cross section is sensitive to the mass of the exchanged object. The result
of a simultaneous fit of the Fermi constant $G_F$ and the propagator mass $M_{ \rm prop}$ to the data is shown 
as a contour in Fig.~\ref{fig:ew_isospin_mass} (right). Fixing the value of  $G_F$ to the world average leads to
$M_{\rm prop}=82.87 \pm 1.87 ({\rm exp}) ^{+0.32}_{-0.18} ({\rm model})$ GeV (H1, see also the line in Fig.~\ref{fig:ew_isospin_mass} right) and $M_{\rm prop}=79.1 \pm 0.77 ({\rm stat.+uncor.}) \pm 0.99 ({\rm corr.\;  syst.})$ GeV (ZEUS). These results are in good agreement with the world average on on-shell mass of $W$-bosons, confirming the picture of CC scattering being due to the $W$ exchange. 

\section{Search for physics beyond the Standard Model}

The HERA collider  is also an  energy frontier machine with a center of mass energy $\sqrt{s}=318$ GeV.
There are two basic approaches to search for physics beyond the SM. In one group of analyses, so called ``generic searches'', one looks for any deviation from the SM prediction. All possible high ${\rm P_T}$ topologies are investigated,
thereby such searches achieve great generality and minimize the probability to miss something interesting. Other analyses perform dedicated searches for predicted signatures of various models for physics beyond the SM.  
\begin{figure}[hhh]
\center
\epsfig{file=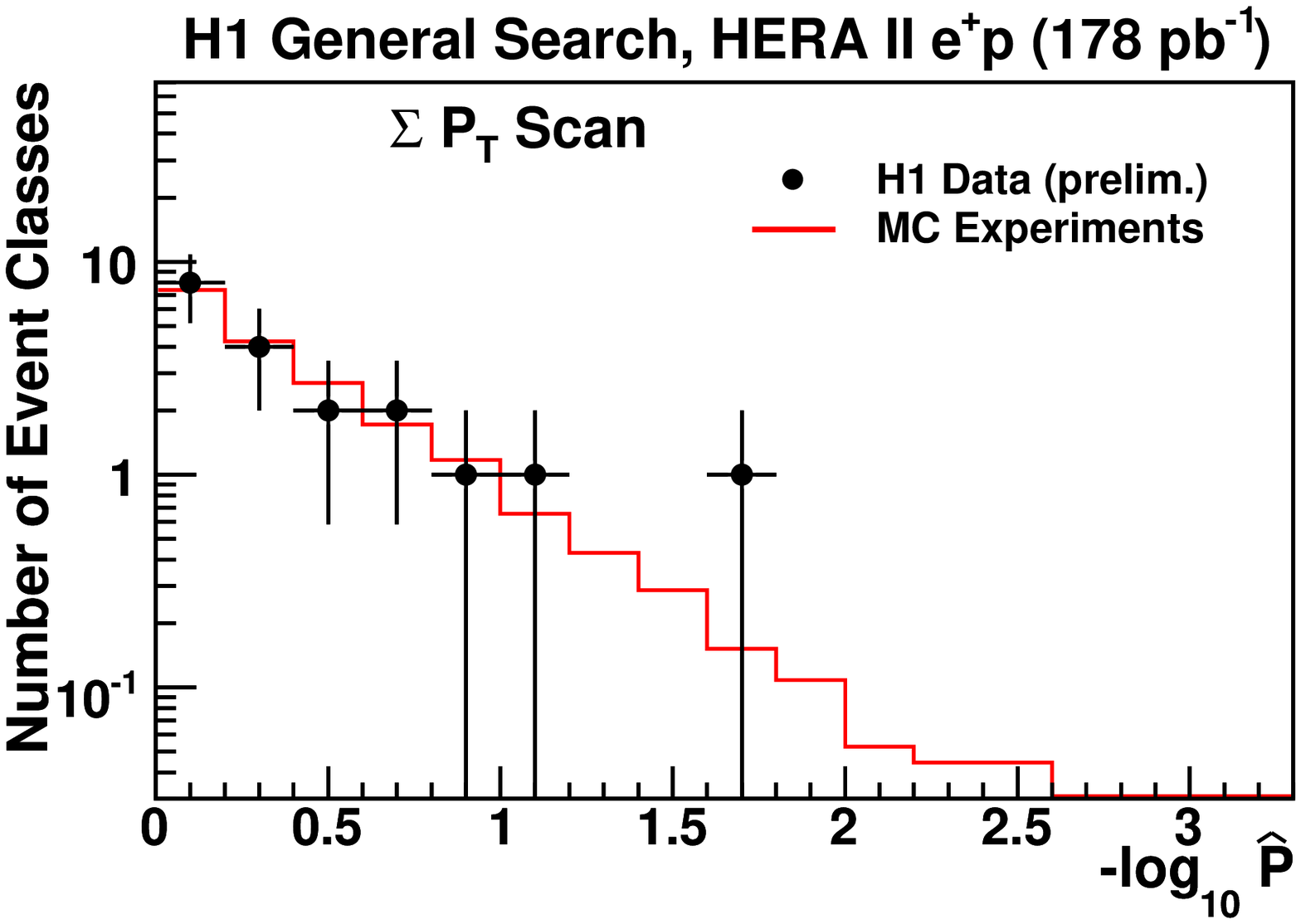,width=.45\textwidth}
\epsfig{file=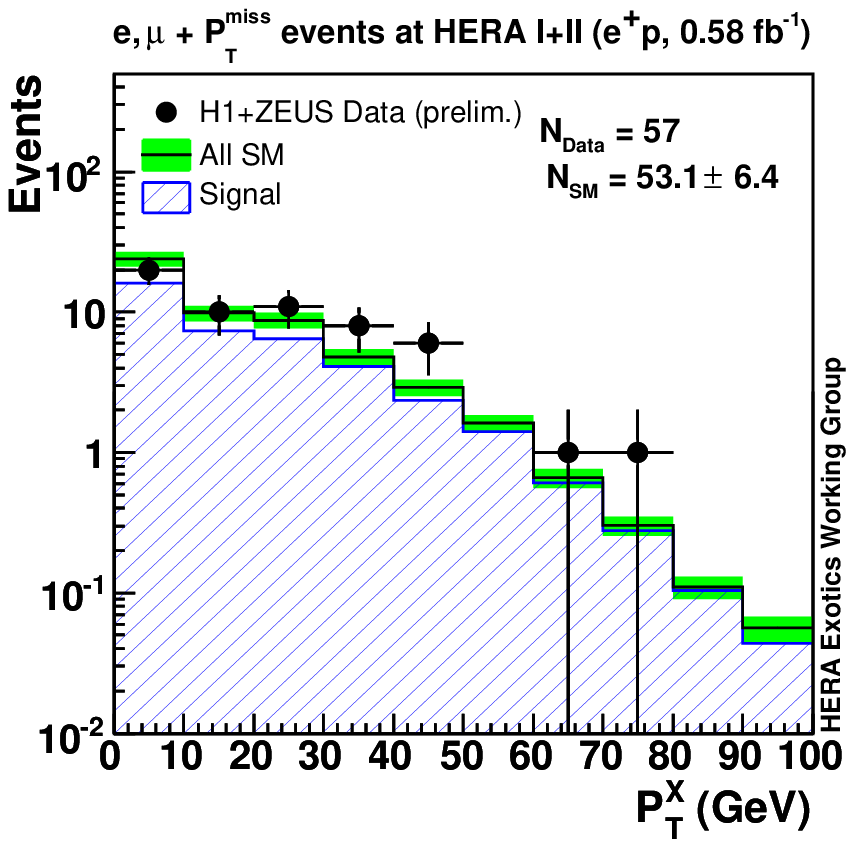,width=.4\textwidth}
\caption{The distributions of likelihoods of deviations from the SM as found in the generic search by the H1 collaboration (left) and the distribution of transverse momentum of hadronic final state ${\rm P_T^X}$  for isolated lepton events as obtained in a combined analysis of H1 and ZEUS data (right). Both distributions are shown for the case of $e^+p$ interactions.}
\label{fig:bsm_generic} 
\end{figure}

The result of an example of generic search for physics beyond the SM can be seen in 
Fig.~\ref{fig:bsm_generic} (left). In this analysis all isolated high ${\rm P_T}$ objects ($e$, $\gamma$, $\mu$, jet, $\nu$) are found in a common phase space defined by ${\rm P_T} > 20$ GeV and scattering angle $10 < \theta < 140^o$. The events, containing at least two high ${\rm P_T}$ objects are classified into exclusive channels, for example electron-jet, jet-jet $\ldots$. The number of events belonging to individual classes  is in good agreement with the SM prediction for all exclusive channels.
While the total number of entries may be in agreement with the SM prediction, possible deviations may be located in particular corners of the phase space. That is why for each exclusive channel, the spectra of invariant
mass and total scalar 
${\rm P_T}$ are constructed, and an automatic procedure is used to search the spectra for the regions with maximal deviation from the SM prediction. To study pure statistical fluctuations of the SM prediction, many MC samples are generated and used to calculate the likelihood of a deviation under the assumption of the SM being valid. The distribution of likelihoods for maximum deviations found in all event classes is shown in Fig.~\ref{fig:bsm_generic} (left). They are in good agreement with expectation. The most deviating event class corresponds to the events containing an electron, jet and neutrino in $e^+ p$ data. 

This class of events, called isolated lepton events, was studied in more detail in a dedicated analysis. It is characterized by high missing transverse momentum (${\rm P_T^{miss}}$), isolated high ${\rm P_T}$ lepton (electron or muon) and a hadronic final state characterized by  its  transverse momentum (${\rm P_T^X}$). The dominant SM process having this signature is the photoproduction of  $W$-bosons with their subsequent decay to a lepton and neutrino. This SM background can be predicted with good accuracy and contributes mainly at rather low  ${\rm P_T^X}$. Analyzing all data from HERA I+II, H1 observes an excess of isolated leptons in the tail of the ${\rm P_T^X}$  distribution of $e^+p$ data. In the high ${\rm P_T^X}$ region  (${\rm P_T^X}>25$ GeV) the significance of the deviation from the SM for $e^+p$ data is on the level of $3 \sigma$ ($21$ observed events vs. $8.8 \pm 1.5$ expected), while the data for $e^-p$  is in agreement with the SM prediction ($3$ observed events vs. $6.9\pm 1$ expected). A similar analysis by ZEUS does not show any excess. The H1 data, combined with the ZEUS data in the phase space of the latter    
are shown in Fig.~\ref{fig:bsm_generic} (right). For the combined analysis, the excess drops to about $2 \sigma$ with $24$ events observed ($14.6\pm 1.9$ expected) in $e^+p$ and 6 events observed ($10.6 \pm 1.4 $ expected) in $e^- p$ collisions.
\begin{figure}[hhh]
\center
\epsfig{file=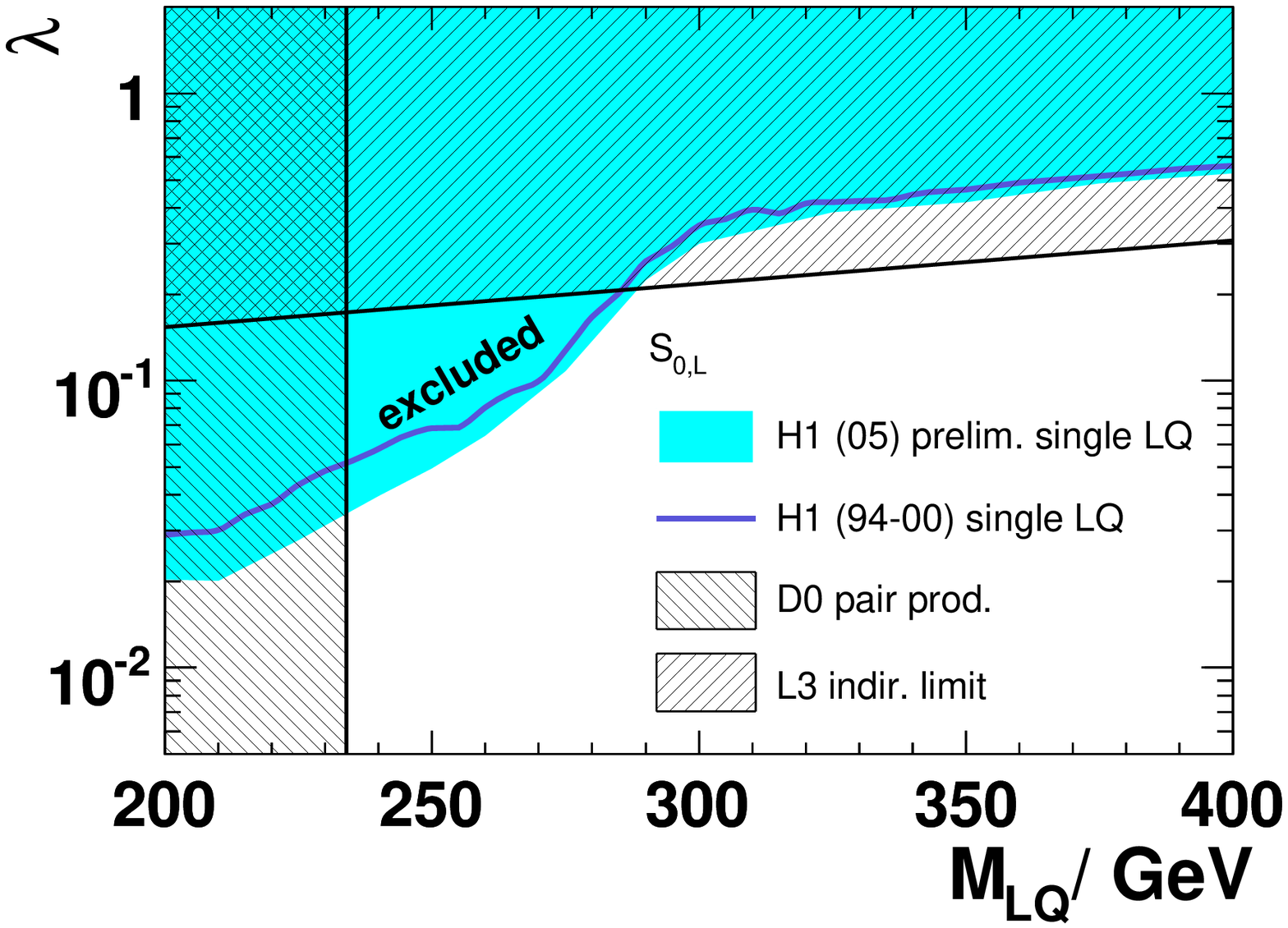,width=.48\textwidth}
\epsfig{file=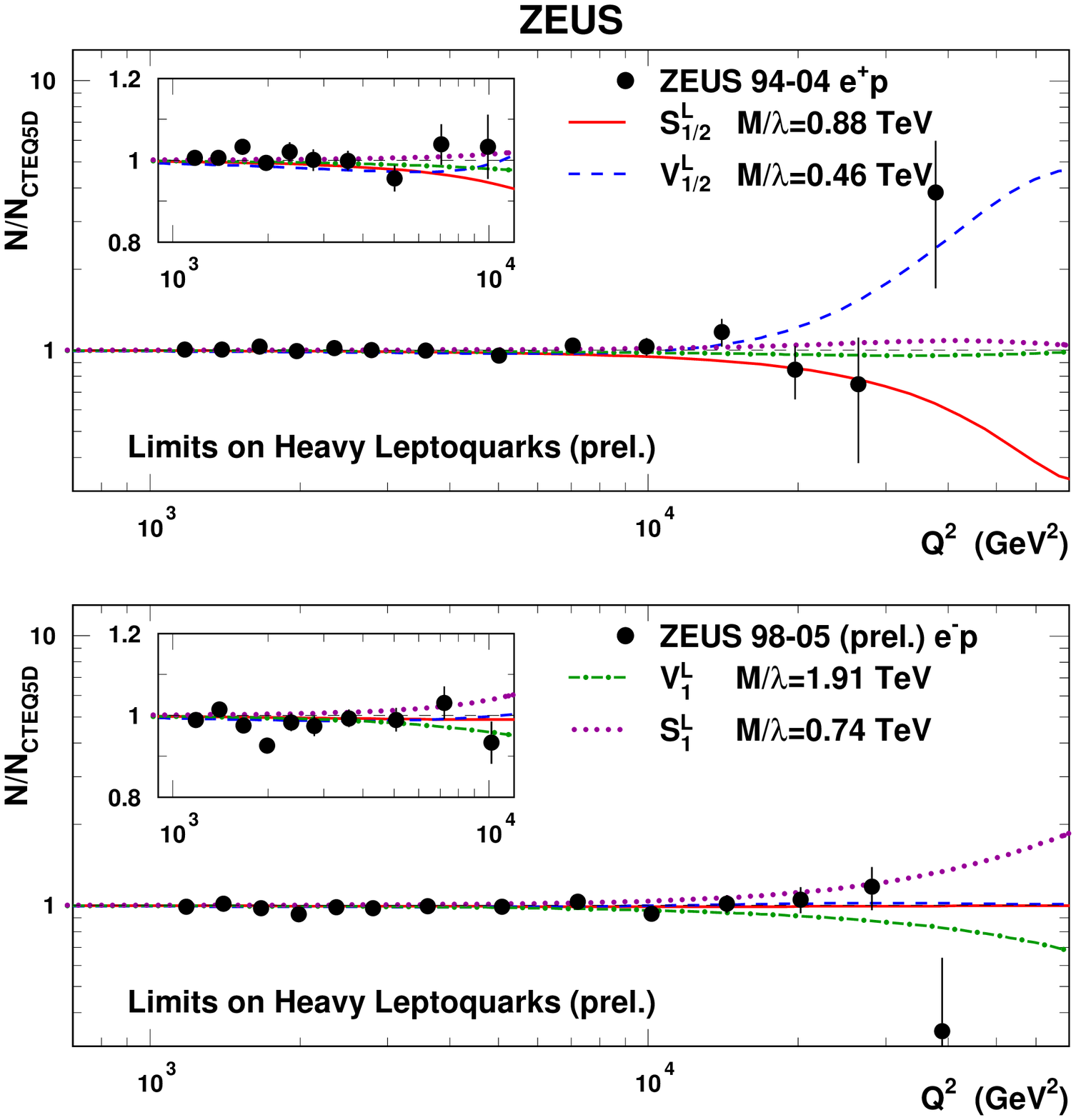,width=.33\textwidth}
\caption{The exclusion limits in the parameter space of leptoquark mass ($M_{LQ}$) and its Yukawa coupling ($\lambda$)  for low mass leptoquarks (left) and the high $Q^2$ NC cross sections used to set confidence limits for high mass leptoquarks (right).  }
\label{fig:bsm_lq} 
\end{figure}
\begin{figure}[hhh]
\center
\epsfig{file=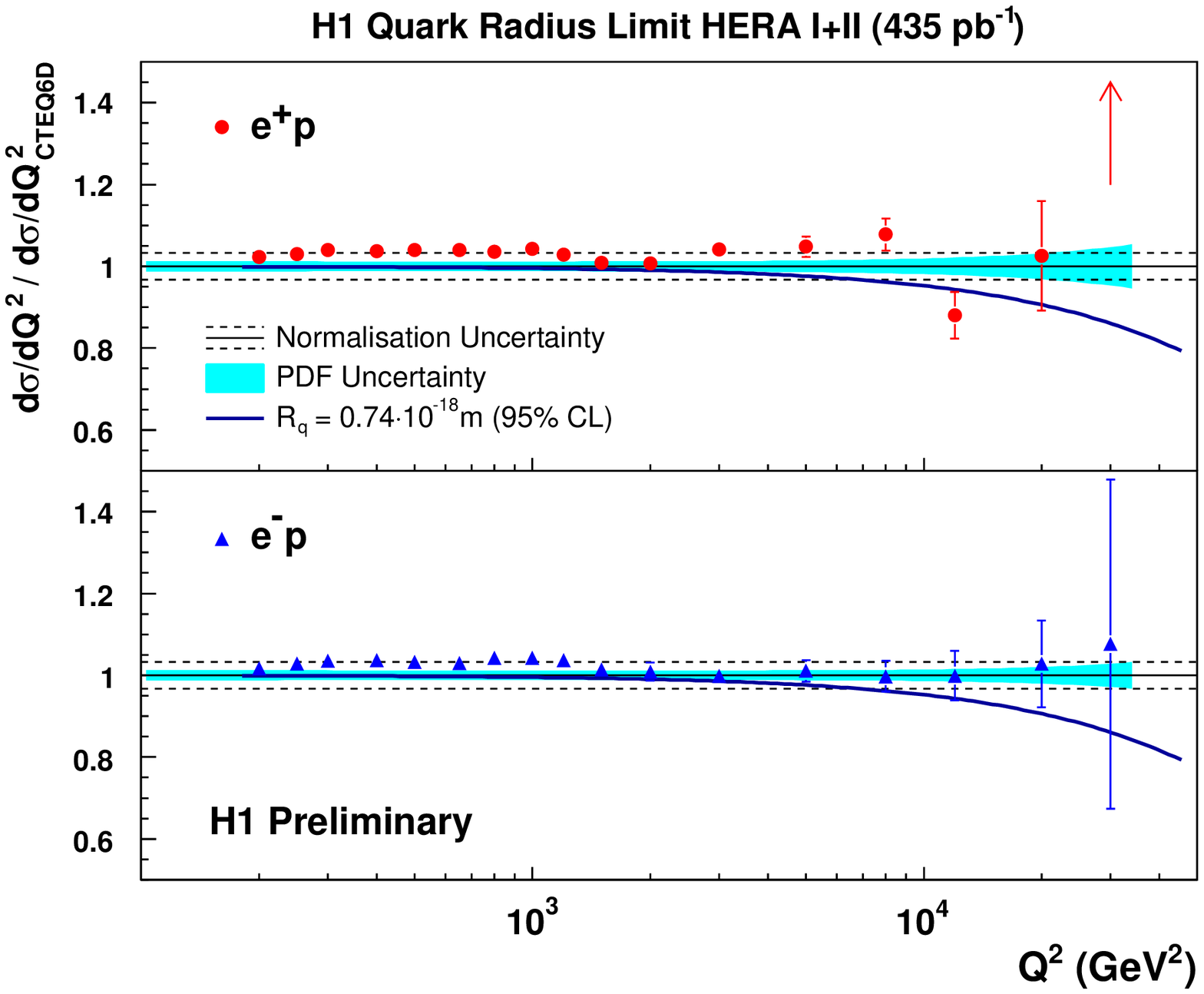,width=.43\textwidth}
\epsfig{file=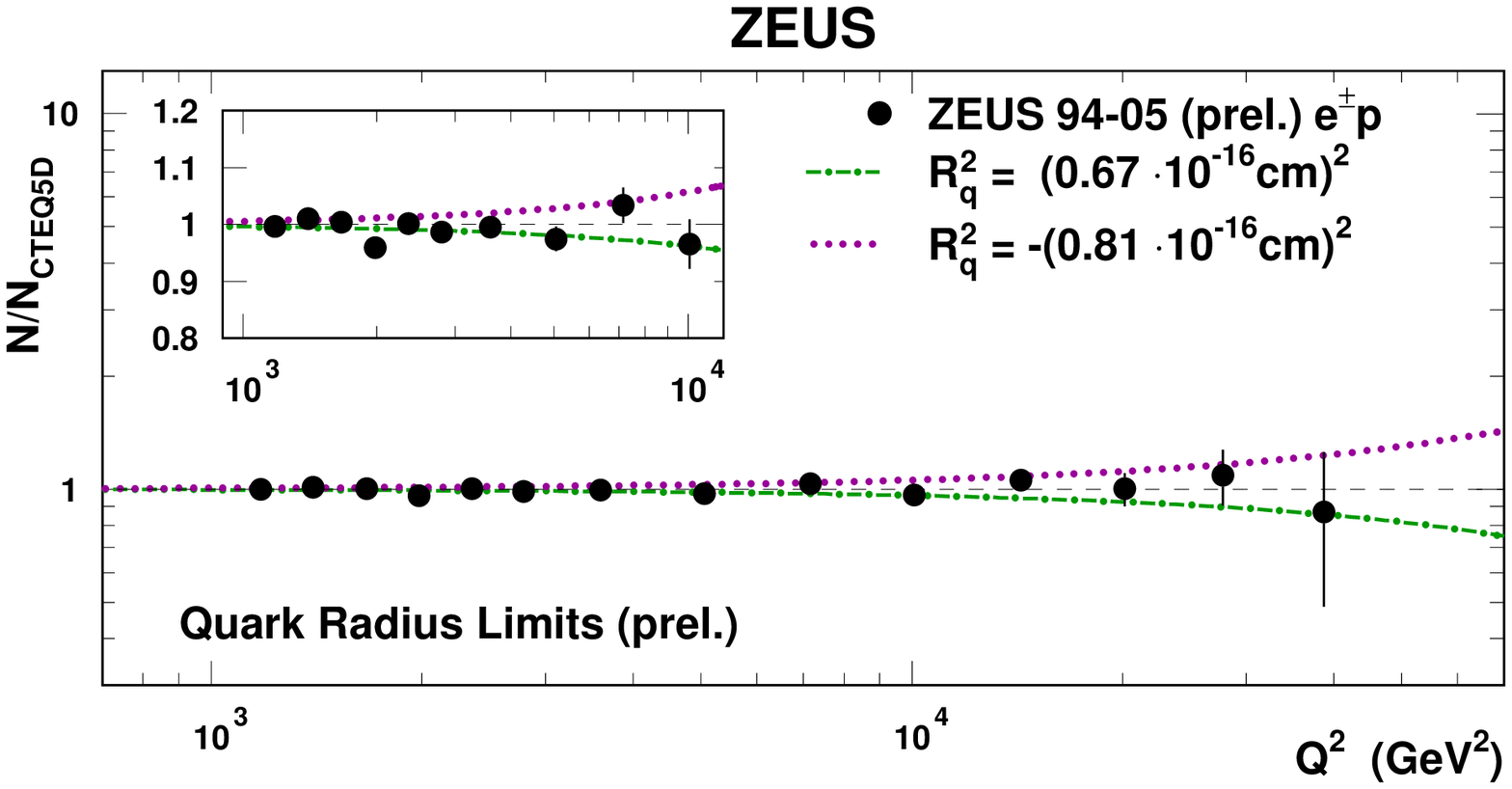,width=.50\textwidth}
\caption{The high $Q^2$ NC cross sections together with the confidence limits on the electromagnetic radius of the quarks. }
\label{fig:bsm_quarkrad} 
\end{figure}

The collisions of $e$ and $p$ beams at HERA are naturally suited to search for leptoquarks, the hypothetical boson particles appearing in various models beyond the SM. Lower mass leptoquarks are expected to be produced in the $s$-channel and  their signature would be a peak in the invariant mass distribution of the lepton-quark system. No such peak is observed, and the data are used to set the exclusion limits on the leptoquark Yukawa coupling $\lambda$ as a function of the leptoquark mass $M_{\rm LQ}$ (see Fig.~\ref{fig:bsm_lq} left).
Leptoquarks with mass larger than $\sqrt{s}$ may still contribute to the NC cross section via  u-channel exchange. In this case, a deviation from the  SM cross section at high $Q^2$ is searched for. As no deviation is observed, the data are used to set limits on the ratio of the leptoquark mass and its Yukawa coupling $M_{\rm LQ}/\lambda$ (see Fig.~\ref{fig:bsm_lq} right).  

The NC cross section at high $Q^2$ is also used to search for a non-pointlike structure of the quarks. A possible structure of quarks would modify the SM cross section at high $Q^2$ according to the formula $d\sigma/dQ^2 = d\sigma^{SM}/dQ^2 f_q(Q^2)$, where $f_q(Q^2)$ is the form-factor of the quark. No deviation from the SM prediction at high $Q^2$ is observed, and the data are used to set confidence limits on the electromagnetic charge radius of the quark, using the dipole approximation for the quark form-factor $f(Q^2) = 1-(<r^2>Q^2)/6$. The results are shown in Fig.~\ref{fig:bsm_quarkrad}, with $95 \%$ confidence limits on the quark radius being $r_q < 0.74 \times 10^{-18}$ m (H1) and $r_q < 0.67 \times 10^{-18}$ m (ZEUS).

\begin{acknowledgments}

I would like to thank R. Pla\v cakyt\. e, G. Grindhammer and M. Klein for usefull discussions and help in preparing this manuscript.

\end{acknowledgments}


\end{document}